**Discrete gap solitons in binary positive-negative index nonlinear waveguide arrays with strong second-order couplings**

Alexander A. Dovgiy*,[1] and Ilya S. Besedin[1,2]

[1]Department of Solid State Physics and Nanosystems, National Research Nuclear University "MEPhI", Kashirskoe shosse 31, 115409 Moscow, Russia

[2]Deutsches Electron-Synchrotron DESY, Notkeshtrasse 85, D-22607 Hamburg, Germany

We report on existence and properties of discrete gap solitons in zigzag arrays of alternating waveguides with positive and negative refractive indices. Zigzag quasi-one-dimensional configuration of waveguide array introduces strong next-to-nearest neighbor interaction in addition to nearest-neighbor coupling. Effective diffraction can be controlled both in size and in sign by the value of the next-to-nearest neighbor coupling coefficient and even can be cancelled. In the regime where instabilities occur, we found different families of discrete solitons bifurcating from gap edges of the linear spectrum. We show that both staggered and unstaggered discrete solitons can become highly localized states near the zero diffraction points even for low powers. Stability analysis has shown that found soliton solutions are stable over a wide range of parameters and can exist in focusing, defocusing and even in alternating focusing-defocusing array.

## I. INTRODUCTION

Diffraction effects in discrete optical systems strongly modify from that ones in homogenous and isotropic media. Peculiarities in diffraction effects appear due to rotational symmetry breaking in discrete optical systems, e.g., in array of waveguides, and canonical laws of diffraction cease to hold. Such systems allow one to control the diffraction either in size or sign by the input conditions (angle of incidence of a beam) [1]. Diffractive beam spreading can even be arrested and diverging light can be focused. Analytical explanation of such phenomena comes from the mathematical relation between longitudinal and transverse wave number components of the wave vector. This relation is analogous to the dispersion relation in the temporal domain and describes the diffraction process in the system considered. In the case of waveguide arrays this relation is strictly periodic; hence, either strength or sign of diffraction depends on the transverse wave number component periodically, which in turn is determined by the tilt of the initial beam. Thus, the light beam can undergo both normal and anomalous diffraction and even can cross the array diffractionless. By using the diffraction properties of waveguide arrays, it is possible to produce structures with reduced, canceled and even reversed diffraction. Results of experiments with such waveguide arrays are presented and compared with the predictions made by coupled-mode theory in [2]. Similar effect was shown to happen in photonic crystals [3].

Diffraction effects play a significant role for the formation of spatial self-localized states (solitons) in nonlinear media. Discrete diffraction, as discussed above, has peculiarities and, as a result, nonlinear response in discrete structures demonstrates novel effects, which have no analogs in continuous systems. In nonlinear waveguide arrays (NOWA) spatial discrete solitons can be formed due to the interplay between discrete diffraction, arising from linear coupling, and waveguide nonlinearity. Due to the possibility of diffraction management in waveguide arrays, different families of discrete solitons can be formed. Thus, both self-focusing and self-defocusing have been achieved experimentally for the same medium, structure (waveguide array), and wavelength [4]. Also, it was predicted analytically by Kivshar [5] that discrete self-focusing may be realized in array of defocusing waveguides when the transverse wave number component of the wave vector lies at the edge of a Brillouin zone. At the base of the Brillouin zone discrete self-focusing occurred to be in nonlinear focusing waveguides in the same wave-guide array [6]. All these listed features of discrete self-localized states (solitons) are consequences of mentioned above diffraction properties of waveguide arrays.

Discussed above studies are dedicated to the analysis of linear and nonlinear properties of uniform waveguide arrays, i.e., arrays composed of equally spaced identical waveguides. However, analysis of nonuniform waveguide arrays (binary ones, arrays with defects, etc.) provides further degrees of freedom. Binary waveguide arrays possess a linear band gap, thus, new kinds of discrete gap solitons can be obtained in such structures [7]. An interesting result is obtained in the case of a binary array with periodical switching of the coupling between successive waveguides (the coupling coefficients differ not only in modulus but also in sign); flat-top and kink solitons can be formed in this structure. Both stationary and "walking" gap solitons moving along the spatial coordinate with a tunable velocity exist for focusing, defocusing and even alternating focusing-defocusing nonlinearity [8]. Efremidis and Christodoulides [9] proposed a zigzag configuration of the waveguide array that can exhibit strong second-order coupling in addition to the nearest-neighbor coupling and this extended coupling affects the lattice dispersion relation within the Brillouin zone. As a result of this band alteration, completely different families of discrete solitons can be obtained in such arrays which are stable over a wide range of parameters. Also, diffraction management is studied in this structure and it can be employed to generate spatial discrete optical solitons at low power levels.

The achievements of modern technologies, i.e. nanotechnologies, allow manufacturing artificial materials with unusual electromagnetic properties, i.e., metamaterials, which possess negative refraction in microwave range [10-15], and more recently in optical range without losses [16]. Compensation of energy losses in metamaterials can be achieved by implantation of components with active molecules or atoms into the structure of these artificial materials. The properties of negative index media can be employed in new various optical components for the integrated or fibre optics. Nonlinear response of such negative index metamaterials (NIM) to electromagnetic waves propagation leads to novel optical phenomena and is studied thoroughly in the last decade [17-19]. In particular, new regimes of nonlinear wave mixing between forward and backward waves can be brought about in NIM [20-28] (for more examples of studies, see the review papers of Ref. [27, 28]). Also, the interface between NIM and positive index media (PIM) presents new features of refraction or localization of electromagnetic waves [29-33]. Interesting examples of mixed PIM-NIM structures providing the forward-backward waves interaction are nonlinear oppositely directed couplers (NODC) [34-36] and waveguide arrays with alternating PIM and NIM waveguides [37-41]. Novel features of nonlinear wave propagation, such as optical bistability [34], slit solitons [35, 38, 40], suppression of modulation instability effect [36, 39, 41] and discrete gap solitons [37] were observed in these structures. In general, such model of PIM-NIM NOWA describing the nonlinear interaction of forward and backward waves in periodic media can be applied to a broad range of metal-dielectric photonic structures, including plasmonic waveguides and metamaterials [42]. Energy localization can be significantly modified by introducing extended interactions (next-to-nearest neighbors) in PIM-NIM NOWA. These extended interactions may appear by exploiting the topological arrangement of the lattice itself. Zigzag geometrical configuration provides necessary deformation of the lattice allowing introducing second-order couplings in the PIM-NIM NOWA and completely different families of discrete soliton solutions can be obtained there in comparison with the ordinary first-order coupling PIM-NIM NOWA. Also, the dispersion relation of waves propagating in zigzag PIM-NIM NOWA contains a band gap due to the alternating sign of refractive index [40, 41] that makes it different from the model considered in Ref. [9]. Thus, zigzag PIM-NIM NOWA provides further degrees of freedom to manipulate energy localization effects, diffraction management and discrete solitons formation.

In a recent study, it was shown that modulation instability effect in zigzag PIM-NIM NOWA disappears regardless of the electromagnetic field power, when the second-order coupling coefficient exceeds a certain threshold, the value of which depends on the transverse wave number component of the wave vector [41]. Thus, a uniform field distribution in the system in question can be both stable and unstable in the same nonlinear media depending on the second-order coupling coefficient and various regimes of nonlinear wave propagation are possible.

*Corresponding author.
E-mail: dovgiyalexandr@gmail.com

In this paper we report about the existence of spatially localized modes for low powers in this quasi-one-dimensional waveguide array including negative index metamaterial channels. We show that the effective diffraction of the array can be controlled both in size and sign, and even can be cancelled under definite values of the second-order coupling coefficient. Zero diffraction points exist both at the base and at the edge of the Brillouin zone. Both staggered and unstaggered bright solitons observed in the system can become highly localized states even at low power levels near these zero diffraction points. We perform stability analysis of these spatially localized states and demonstrate their stability over the wide range of parameters. We present new example of periodic photonic structure allowing describing nonlinear interaction of forward and backward waves.

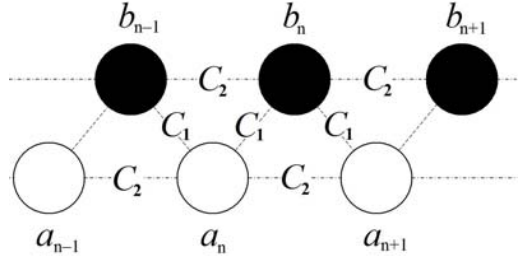

FIG. 1. Zigzag PIM-NIM NOWA (cross section). Empty circles indicate PIM waveguides; filled ones – NIM waveguides.

## II. PHYSICAL MODEL

The coupling between second-order neighbors can be controlled by an angle between the lines connecting neighboring waveguides [9]. When the value of this angle is equal to $\pi$ the second-order interactions are extremely weak and the system is reduced to a first-order interacting waveguide array. Reduction of this angle leads to the increase of second-order coupling coefficient and the waveguide array takes a zigzag configuration (Fig. 1). Thus, zigzag NOWA is the more general case of waveguide arrays which allows considering extended interactions (beyond the nearest neighbors). In this case, stationary field distribution in the physical system in question is described by following nonlinear discrete differential equations [41]:

$$\begin{cases} i\partial_z a_n + \omega a_n + C_1(b_{n-1} + b_n) + C_2(a_{n-1} + a_{n+1}) + \chi_1 |a_n|^2 a_n = 0, \\ -i\partial_z b_n + \omega b_n + C_1(a_n + a_{n+1}) + C_2(b_{n-1} + b_{n+1}) + \chi_2 |b_n|^2 b_n = 0, \end{cases} \quad (1)$$

where $a_n$ and $b_n$ are normalized field amplitudes in PIM and NIM waveguides, respectively, and $n = 0, \pm 1, \pm 2, ...$ is a number of a coupler in the array (see Fig. 1), $z$ is the propagation distance. $C_1$ and $C_2$ are first-order (nearest neighbors interaction) and second-order (next to nearest neighbors interaction) coupling coefficients, $\chi_1$ and $\chi_2$ are normalized nonlinear susceptibilities of PIM and NIM waveguides, respectively, and $\omega > 0$ is a mismatch between propagation constants. Due to opposite signs of Poynting vectors in PIM and NIM waveguides, that provides an effective feedback mechanism in the system under consideration, there is a minus sign in front of the spatial derivative in the second equation of the system (1) unlike the first equation. Therefore, the physical inputs for the field amplitudes $a_n$ and $b_n$ are positioned on $z = 0$ and $z = L$, respectively, where $L$ is the length of the array along $z$ direction. But this boundary problem of the array in question can be simplified by considering stationary field distributions in wave-guides [37] and, therefore, both fields can be given at $z = 0$ (i.e. $b_n$ are given at its physical output).

*Corresponding author.
E-mail: dovgiyalexandr@gmail.com

We can determine the total power in the array as

$$P = \sum_n P_n = \sum_n \left(|a_n|^2 + |b_n|^2\right),\qquad(2)$$

as well as the Hamiltonian

$$H = 2\text{Re}\sum_n \left\{ C_1\left(a_n b_n^* + a_n b_{n-1}^*\right) + C_2\left(a_n a_{n-1}^* + b_n b_{n-1}^*\right) + \tfrac{1}{4}\left(\chi_1 |a_n|^4 + \chi_2 |b_n|^4\right)\right\} + \omega P \qquad(3)$$

from where one can obtain the equations of motion [Eq. (1)] via $i\dot{a}_n = -\partial H/\partial a_n^*$ and $i\dot{b}_n = \partial H/\partial b_n^*$ (an overdot stands for a derivative with respect to $z$). It's important to note that the Hamiltonian (3) and the total power (2) are conserved quantities of Eq. (1).

In order to derive equations of motion in the so-called continuous approximation (i. e., slowly varying amplitudes), we consider more general case of Eq. (1) describing nonstationary field distribution in the linear array [41]:

$$\begin{cases} i(\partial_z + \partial_t)a_n + \nu a_n + C_1(b_{n-1} + b_n) + C_2(a_{n-1} + a_{n+1}) = 0, \\ i(-\partial_z + \partial_t)b_n + \nu b_n + C_1(a_n + a_{n+1}) + C_2(b_{n-1} + b_{n+1}) = 0, \end{cases}\qquad(4)$$

where $\nu$ is a phase mismatch between adjacent waveguides. If we make the Fourier transform in such a way as $a_n = \sum_{k,q} u_{k,q} \exp i(kz + qn - \omega_{k,q} t)$ and $b_n = \sum_{k,q} v_{k,q} \exp i(kz + qn + q/2 - \omega_{k,q} t)$, it's easy to obtain the following equation for the Fourier amplitudes: $\hat{L}_{k,q} x_{k,q} = 0$, where $x_{k,q} = (u_{k,q}\ v_{k,q})^T$ and

$$\hat{L}_{k,q} = \begin{pmatrix} \omega - k + 2C_2 \cos q & 2C_1 \cos(q/2) \\ 2C_1 \cos(q/2) & \omega + k + 2C_2 \cos q \end{pmatrix},\qquad(5)$$

$\omega = \omega_{k,q} + \nu$, from where one can easily obtain the dispersion relation of system (4) via equation $\det \hat{L}_{k,q} = 0$, which determines the frequency $\omega_{k,q}$ as a function of $k$ and $q$:

$$\omega = -2C_2 \cos q \pm \sqrt{k^2 + 4C_1^2 \cos^2 q/2}.\qquad(6)$$

We can introduce the slowly varying amplitudes as follows:

$$a_n = \varphi(z,t,n)\exp i(k_0 z + q_0 n - \omega_0 t),\ b_n = \psi(z,t,n)\exp i(k_0 z + q_0 n - \omega_0 t),\qquad(7)$$

where $\varphi$ and $\psi$ are slowly varying functions, $k_0$, $q_0$ and $\omega_0 = \omega_{k_0,q_0}$ are spatial and temporal carrier frequencies, respectively. The frequencies of quasi-monochromatic envelopes with a narrow spectral width slightly deviate from the carriers. From Eq. (7) it follows that the Fourier images of slowly varying amplitudes satisfy the relations $\chi_{\tilde{k},\tilde{q}} = x_{k_0+\tilde{k},q_0+\tilde{q}}$ and $\hat{L}_{k_0+\tilde{k},q_0+\tilde{q}} \chi_{\tilde{k},\tilde{q}} = 0$, where $\chi_{\tilde{k},\tilde{q}} = (\varphi_{\tilde{k},\tilde{q}}\ \psi_{\tilde{k},\tilde{q}})^T$ and $\tilde{k}$, $\tilde{q}$ are small deviations from the carrier frequencies. Hence, we have the following Taylor series expansion:

$$\left(\hat{L}_{k_0,q_0} + \tilde{k}\left.\frac{\partial \hat{L}}{\partial \tilde{k}}\right|_{\tilde{k}=0} + \tilde{q}\left.\frac{\partial \hat{L}}{\partial \tilde{q}}\right|_{\tilde{q}=0} + \frac{\tilde{q}^2}{2}\left.\frac{\partial^2 \hat{L}}{\partial \tilde{q}^2}\right|_{\tilde{q}=0} + ...\right)\chi_{\tilde{k},\tilde{q}} = 0.\qquad(8)$$

Proceeding up to the fourth-order term in the Taylor series one can obtain equations for the slowly varying amplitudes $\varphi$ and $\psi$ by applying the inverse Fourier transform to the Eq. (8):

*Corresponding author.
E-mail: dovgiyalexandr@gmail.com

$$\begin{cases} i\varphi_z + i\varphi_t + d_1^{(2)}\psi + iv_g^{(2)}\psi_x + d_2^{(2)}\psi_{2x} + id_3^{(2)}\psi_{3x} + d_4^{(2)}\psi_{4x} + d_1^{(1)}\varphi + iv_g^{(1)}\varphi_x + d_2^{(1)}\varphi_{2x} + id_3^{(1)}\varphi_{3x} \\ + d_4^{(1)}\varphi_{4x} = 0, \\ -i\psi_z + i\psi_t + d_1^{(2)}\varphi + iv_g^{(2)}\varphi_x + d_2^{(2)}\varphi_{2x} + id_3^{(2)}\varphi_{3x} + d_4^{(2)}\varphi_{4x} + d_1^{(1)}\psi + iv_g^{(1)}\psi_x + d_2^{(1)}\psi_{2x} + id_3^{(1)}\psi_{3x} \\ + d_4^{(1)}\psi_{4x} = 0, \end{cases} \quad (9)$$

where $x = s/l$ is a normalized coordinate, $l$ is the distance between adjacent waveguides, $s$ is the actual transverse skew coordinate along the zigzag path $(n \to s/l)$, and the coefficients $d_1^{(1)} = 2C_2 \cos q_0$, $d_1^{(2)} = 2C_1 \cos(q_0/2)$,

$$v_g^{(1)} = 2C_2 \sin q_0, \quad v_g^{(2)} = C_1 \sin(q_0/2), \quad (10)$$

$$d_2^{(1)} = C_2 \cos q_0, \quad d_2^{(2)} = \frac{C_1}{4} \cos(q_0/2), \quad (11)$$

$$d_3^{(1)} = \frac{C_2}{3} \sin q_0, \quad d_3^{(2)} = \frac{C_1}{24} \sin(q_0/2), \quad (12)$$

$$d_4^{(1)} = \frac{C_2}{12} \cos q_0, \quad d_4^{(2)} = \frac{C_1}{192} \cos(q_0/2). \quad (13)$$

For the stationary field distribution in the nonlinear array in question Eq. (9) can be written as

$$\begin{cases} i\varphi_z + \omega\varphi + d_1^{(2)}\psi + iv_g^{(2)}\psi_x + d_2^{(2)}\psi_{2x} + id_3^{(2)}\psi_{3x} + d_4^{(2)}\psi_{4x} + d_1^{(1)}\varphi + iv_g^{(1)}\varphi_x + d_2^{(1)}\varphi_{2x} \\ + id_3^{(1)}\varphi_{3x} + d_4^{(1)}\varphi_{4x} + \chi_1 |\varphi|^2 \varphi = 0, \\ -i\psi_z + \omega\psi + d_1^{(2)}\varphi + iv_g^{(2)}\varphi_x + d_2^{(2)}\varphi_{2x} + id_3^{(2)}\varphi_{3x} + d_4^{(2)}\varphi_{4x} + d_1^{(1)}\psi + iv_g^{(1)}\psi_x + d_2^{(1)}\psi_{2x} \\ + id_3^{(1)}\psi_{3x} + d_4^{(1)}\psi_{4x} + \chi_2 |\psi|^2 \psi = 0. \end{cases} \quad (14)$$

The latter equations (14) are the so-called continuous approximation of the discrete differential equations (1) when the field amplitudes are slowly varying functions with respect to waveguide's number. The coefficients $v_g^{(1)}$ and $v_g^{(2)}$ can be attributed to the wave's spatial group velocity, and $d_j^{(i)}$ ($i = 1, 2; j = 2, 3, 4$) represent the second-, third- and fourth-order diffraction effects in the array, respectively.

### III. DIFFRACTION RELATION AND MODULATION INSTABILITY

In Eq. (6) $k$ and $q$ are the longitudinal and the transverse wave number components of the wave vector, respectively, and they are independent variables of the frequency function $\omega(k,q)$, that describes the dispersion in the array. To study the stationary field problem the frequency should be fixed, hence in this case Eq. (6) determines an implicit function $k(q)$, that can be expressed as

$$k^2(q) = (\omega + 2C_2 \cos q)^2 - 4C_1^2 \cos^2(q/2). \quad (15)$$

The latter expression (15) determines the so-called diffraction relation and it can be used to analyze the diffraction management in the array under consideration. Fig. 2 depicts diffraction curve in the domain of the first Brillouin zone for different values of second-order coupling coefficient which in turn affects sufficiently at the behavior of diffraction and as a consequence at the bright solitons formation. As it is said above, the second-order interactions in the array can be controlled by the angle between the lines connecting neighboring waveguides and the system can be reduced to the first-order interacting array. In this case the second-order coupling coefficient is equal to zero and the diffraction relation takes the form of that one considered in ordinary PIM-NIM array [37]. It should be noted that we will discuss only the upper branch of the diffraction curve. For the lower branch all discussions are exactly reversed. When $q$ lies in the range $-\pi/2 < q < \pi/2$ the curvature of the diffraction relation is positive [$k''(q) > 0$] (Fig. 2a) and it has the form of a discrete Schrödinger type (DS-type) diffraction [1, 2, 4, 9], but reversed. In this region of $q$ the effective diffraction of the array is "anomalous" and bright solitons are expected to arise in defocusing waveguides at the base of the Brillouin zone with eigenvalues lying in the internal finite band gap. No bright solitons were observed in focusing array in this case. However, when PIM and NIM waveguides have nonlinearities of different signs, e.g. focusing and defocusing, respectively, it was observed that families of finite gap solitons bifurcate only from a bottom of the band gap in ordinary PIM-NIM array despite the fact that the finite gap edges are symmetric [37]. The nonlinearity breaks inversion symmetry in the reciprocal space. Contrariwise, in the regions $\pi/2 < |q| < \pi$ the curvature is negative [$k''(q) < 0$] and the effective diffraction of the array is "normal" and of the DS-type in this case. Therefore self-localization can become possible now in focusing waveguides and bright soliton solutions may occur at the edge of the Brillouin zone with eigenvalues lying in the external semi-infinite band gap. In Ref. [37] it was found that there exists more than one family of symmetric and antisymmetric solitons bifurcating from the gap edges of the linear spectrum.

*Corresponding author.
E-mail: dovgiyalexandr@gmail.com

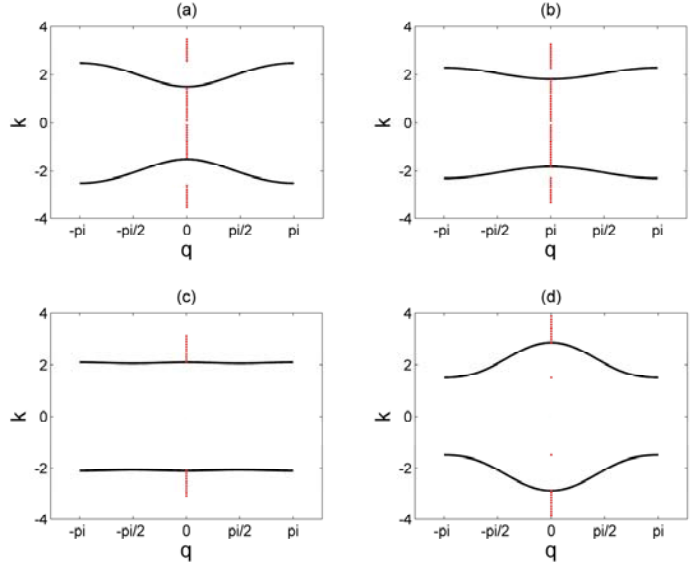

FIG. 2. Diffraction curves (solid lines) for $C_2 = 0, 0.1, 0.2, 0.5$ shown in (a), (b), (c), (d), respectively, when $C_1 = 1$, $\omega = 2.5$. Dots correspond to bright soliton solutions' eigenvalues lying in the band gaps.

    The increase of the second-order coupling coefficient leads to the finite band gap expansion and the diffraction curve becomes broader for both the base and edge of the Brillouin zone (Fig. 2b). As a result, the bright soliton solutions become narrower (occupying less amount of lattice sites) in comparison with those ones with smaller values of $C_2$ for the same power $P$. As can be seen from the Fig. 2d the curvature of the diffraction curve changes its sign, i.e. it becomes negative ("normal" diffraction) in the region $-\pi/2 < q < \pi/2$ and positive ("anomalous" diffraction) in the regions $\pi/2 < |q| < \pi$ in comparison with Fig. 2a,b. Thus, the effective diffraction of the array becomes like that one in the DS model when $C_2$ becomes commensurate with $C_1$. The bright solitons were observed only at the base of the Brillouin zone with eigenvalues lying in the external semi-infinite band gaps when the waveguides of the array in question are focusing. It's important to note that so-called zero diffraction (zd) points [$k''(q) = 0$] exist there for both the base and edge of the Brillouin zone. To determine these zero diffraction points and appropriate values of $C_2$ we can use Taylor series expansion of the diffraction relation (15) for both the base and edge of the Brillouin zone: $k^2 = (k^{(0)}_{q=0,\pi})^2 + 2\sum_{m=1}^{} (-1)^m \kappa^{(m)}_{q=0,\pi} q^{2m} / (2m)!$, where

*Corresponding author.
E-mail: dovgiyalexandr@gmail.com

$\kappa_{q=0}^{(m)} = 2^{2m}C_2^2 + 2\omega C_2 - C_1^2$ and $\kappa_{q=\pi}^{(m)} = 2^{2m}C_2^2 - 2\omega C_2 + C_1^2$ are $(m+1)$th-order diffraction coefficients, $k_{q=0}^{(0)} = \pm[(\omega + 2C_2)^2 - 4C_1^2]$ and $k_{q=\pi}^{(0)} = \pm(\omega - 2C_2)$ are band edges for the base and for the edge of the Brillouin zone, respectively. Here, one can easily obtain the values of $C_2$ corresponding to zero diffraction points of $(m+1)$th-order from the condition $\kappa_{q=0,\pi}^{(m)} = 0$: $C_{2q=0}^{(m)zd} = (\sqrt{(\omega/2^m)^2 + C_1^2} - \omega/2^m)/2^m$ - for the base and $C_{2q=\pi}^{(m)zd} = (\omega/2^m \pm \sqrt{(\omega/2^m)^2 - C_1^2})/2^m$ - for the edge of the Brillouin zone. For the values of parameters used in Fig. 2 ($C_1 = 1$, $\omega = 2.5$) we can estimate the value of $C_2$ in which the second-order effective diffraction in the array disappears: $C_{2q=0}^{(1)zd} \approx 0.175$ and $C_{2q=\pi}^{(1)zd} = 0.25, 1$. Hence, discrete solitons near zero diffraction points can be observed both for the base and edge of the Brillouin zone in contrast to ordinary zigzag array studied in [9] where such nonlinear states was observed only at the edge of the Brillouin zone. This class of solutions with eigenvalues positioned deep inside the band gap represent highly localized states occupying, in essence, 1–3 lattice sites. As the value of $C_2$ increases above 0.25 the diffraction curve becomes narrower and, as a result, the bright soliton solutions become broader. Thus, the array under consideration provides further degrees of freedom for diffraction management and more ways to generate spatial discrete optical solitons at low power levels in comparison with the arrays considered in [9, 37].

To not look unfounded, we investigate the modulation instability (MI) of the plane wave solution of Eq. (1) $a_n = a\exp i(kz + qn)$ and $b_n = b\exp i(kz + qn + q/2)$ with respect to small perturbations. The plane wave solution's amplitudes are coupled by the following equation $2C_1\cos(q/2)b = (k(q) - \omega - 2C_2\cos q)a$ and $k(q)$ is determined by Eq. (15). As it is known, the instability of perturbed continuous waves is closely related with the presence of spatial bright solitons and occurs in the system due to the interplay between nonlinear interaction and diffraction effects. Therefore, the presence of MI can be considered as a precursor to bright soliton formation. We investigate the linear stability by perturbing the amplitude and the phase of the plane wave solution as $a_n = (a + A_n)\exp i(kz + qn + \Phi_n)$ and $b_n = (b + B_n)\exp i(kz + qn + q/2 + \Psi_n)$, where $A_n(z), B_n(z)$ and the differences $\Phi_n(z) - \Psi_n(z)$ are assumed to be small in comparison with the parameters of the plane wave solution. After the linearization of Eq. (1) in these small perturbations and by applying the transformation for these quantities as $(A_n, \Phi_n) \equiv (A, \Phi)\exp i(Kz + Qn)$ and $(B_n, \Psi_n) \equiv (B, \Psi)\exp i(Kz + Qn + Q/2)$ we obtain the following equation between the longitudinal $K$ and the transverse $Q$ wave number components of the perturbations' wave vector: $\hat{S}_Q \vec{g} = K\hat{E}\vec{g}$, where $\vec{g} = (A, \Phi, B, \Psi)^T$, $\hat{E}$ is a $4 \times 4$ unit matrix and $\hat{S}_Q = \{s_{ij}\}$ $(i, j = 1,...,4)$ is referred to as stability matrix which is used to investigate MI in the system in question. The coefficients of this $4 \times 4$ stability matrix are given by:

$$s_{11} = -s_{22} = s_{33} = s_{44} = 2C_2 \sin q \sin Q,$$
$$s_{12} = -2i(C_1 b\cos(q/2) + 2C_2 a\cos q \sin^2(Q/2)),$$
$$-s_{13} = s_{31} = (-a/b)s_{24} = (b/a)s_{42} = 2C_1\sin(q/2)\sin(Q/2),$$
$$(1/b)s_{14} = bs_{41} = (-a)s_{23} = (-1/a)s_{32} = 2iC_1\cos(q/2)\cos(Q/2),$$
$$s_{21} = -i(3\chi_1 a^2 + \omega - k(q) + 2C_2 \cos q \cos Q)/a,$$
$$s_{34} = 2i(C_1 a\cos(q/2) + 2C_2 b\cos q \sin^2(Q/2)),$$
$$s_{43} = i(3\chi_2 b^2 + \omega + k(q) + 2C_2 \cos q \cos Q)/b.$$

*Corresponding author.
E-mail: dovgiyalexandr@gmail.com

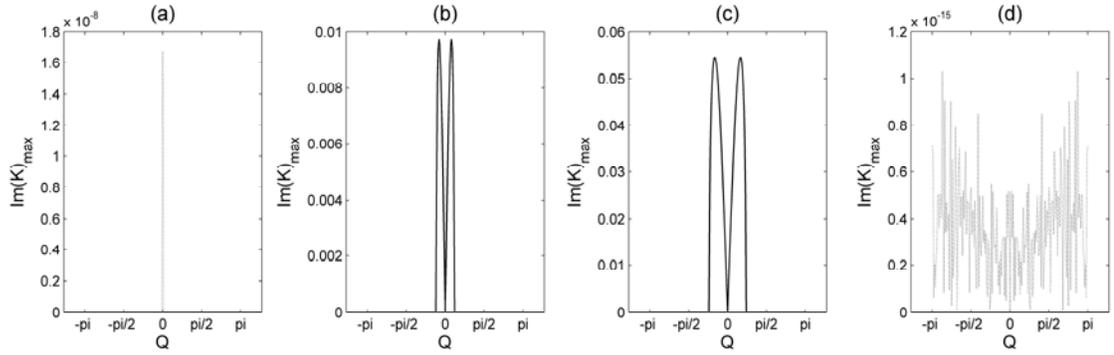

FIG. 3. Modulation instability growth rate (maximum imaginary part of the stability matrix's eigenvalues) for $C_2 = 0.175, 0.176, 0.24, 0.25$ shown in (a), (b), (c), (d), respectively, when $a = 1$, $\chi_1 = \chi_2 = 1$, $C_1 = 1$, $\omega = 2.5$. Figures (a), (b) correspond to $q = 0$ while (c) and (d) to $q \cong \pi$.

MI occurs when at least one of the stability matrix's eigenvalues possesses a nonzero imaginary part which results in an exponential growth of the phase and the amplitude of plane wave solution with the perturbations. To find the eigenvalues of $\hat{S}_Q$ one should solve the diagonalization problem $\det\|\hat{S}_Q - K\hat{E}\| = 0$. We made numerical diagonalization of the stability matrix and determined the regions where the plane wave solution is stable or unstable as a function of $q$ and $C_2$ in focusing array under consideration. Herein, MI totally absent at the base of the Brillouin zone ($q = 0$) when $0 \leq C_2 \leq C_{2q=0}^{(1)zd} \approx 0.175$ [Fig. 3(a)]. Otherwise, the plane wave solution occurs to be unstable when $C_2 > C_{2q=0}^{(1)zd}$ [Fig. 3(b)]. At the ambit of the edge of the Brillouin zone ($q \cong \pi$) MI occurs when $0 \leq C_2 < C_{2q=\pi}^{(1)zd} = 0.25$ [Fig. 3(c)] and totally absent when $C_{2q=\pi}^{(1)zd} \leq C_2 \leq 1$ [Fig. 3(d)]. These results are in full agreement with the conclusions previously drawn from the linear diffraction diagram.

## IV. DISCRETE SOLITONS

In this section we will investigate the existence and stability of bright soliton solutions that are self-localized states in the transverse discrete direction of the array in question. In the continuous approximation these solitons can be described by a nonlinear Schrödinger-type equation [Eq. (14) when $d_3^{(1,2)}$ and $d_4^{(1,2)}$ are negligible], but only for broad enough beams with narrow spectral width. If we want to examine the existence and properties of highly localized states with a good precision, it is necessary to make numerical analysis of discrete Eq. (1). As it is discussed above the discrete solitons can bifurcate in the band gaps from gap edges for the regions where the necessary balance between the diffraction and nonlinear interaction takes place. In order to make analytical investigation we will perform standard multiple-scale expansion procedure as it was performed in [37]. So, we are interested in discrete soliton solutions of Eq. (1) having the form $a_n(z) = u_n \exp i(\kappa z + qn)$, $b_n(z) = v_n \exp i(\kappa z + qn)$, where $u_n$ and $v_n$ are real and vanish as $n \to \pm\infty$, $\kappa$ is the solution's eigenvalue, and two cases of particular interest – unstaggered and staggered solutions, i.e. $q = 0$ and $\pi$, respectively, will be treated separately in the rest of this section.

### A. At the base of the Brillouin zone: q=0

In this case Eq. (1) is reduced to the following system of nonlinear algebraic equations:

*Corresponding author.
E-mail: dovgiyalexandr@gmail.com

$$\begin{cases} (\omega-\kappa)u_n + C_1(v_{n-1}+v_n) + C_2(u_{n-1}+u_{n+1}) + \chi_1|u_n|^2 u_n = 0, \\ (\omega+\kappa)v_n + C_1(u_n+u_{n+1}) + C_2(v_{n-1}+v_{n+1}) + \chi_2|v_n|^2 v_n = 0. \end{cases} \quad (18)$$

We are interested in solutions with exponential decay at $n \to -\infty$ ($n \to +\infty$), hence, we can require the relation $u_{n+1}/u_n = v_n/v_{n-1} = \alpha$ ($u_n/u_{n+1} = v_{n-1}/v_n = \alpha$) to hold, where $\alpha$ is real and $|\alpha|>1$. For exponentially decaying solutions the nonlinear terms of Eq. (18) can be neglected for large $n$, and by substituting the latter expression to Eq. (18) one can easily obtain the relation which establishes domains of the solution's eigenvalue $\kappa$, for which solitons exist: $[(\omega-\kappa)\alpha + C_2(1+\alpha^2)][(\omega+\kappa)\alpha + C_2(1+\alpha^2)] = C_1^2\alpha(1+\alpha)^2$. The results of numerical analysis of this fourth order algebraic equation are depicted in Fig. 2 with dots. In order to satisfy the condition $|\alpha|>1$ the soliton solution's eigenvalues should be in the band gaps where the light propagation is forbidden [Fig. 2(a, b)]. For some values of the second-order coupling coefficient $C_2$ the internal finite band gap domain of the soliton solution's eigenvalue may be degenerate [Fig. 2(c, d)].

To consider the bifurcation of solitons from gap edges we will shift the soliton solution's eigenvalue towards the gap: $\kappa = k_{q=0}^{(0)} + \kappa_2\varepsilon^2$, where $k_{q=0}^{(0)}$ indicate gap edges as were determined in the previous section, $\varepsilon \ll 1$ is a small parameter characterizing the shift of the eigenvalue towards the gap and the sign of $\kappa_2$ determines the direction of the shift. Performing standard multiple-scale series expansion $u_n = \varepsilon \sum_{m=0} \varepsilon^m U_m(x)$ and $v_n = \varepsilon \sum_{m=0} \varepsilon^m V_m(x)$, where $x = \varepsilon n$, we obtain the relation $V_0 = \beta U_0$, where $\beta = [k_{q=0}^{(0)} - (\omega + 2C_2)]/2C_1 < 0$. Proceeding up to the third-order term in the multiple-scale series we obtain the following stationary nonlinear Schrödinger (NLS) equation:

$$\gamma \frac{d^2 U_0}{dx^2} - \kappa_2(1-\beta^2)U_0 + (\chi_1 + \beta^4 \chi_2)U_0^3 = 0, \quad (19)$$

where $\gamma = \beta C_1/2 + C_2(1+\beta^2)$ can be interpreted as the second-order diffraction coefficient. Note that $\gamma$ is equal to zero when $C_2 = C_{2q=0}^{(1)zd}$. Eq. (19) has a well-known bright soliton solution

$$U_0 = A/\cosh(bx), \quad (20)$$

where $A = [2\kappa_2(1-\beta^2)/(\chi_1+\beta^4\chi_2)]^{1/2}$ and $b = [\kappa_2(1-\beta^2)/\gamma]^{1/2}$. If $C_2 < C_{2q=0}^{(1)zd}$ then $\gamma$ is negative and, as it's seen from the latter expressions for $A$ and $b$, the shift of the soliton's eigenvalue should be towards the internal finite band gap and the quantity $(\chi_1+\beta^4\chi_2)$ should be negative, hence, no bright solitons can bifurcate from the gap edges when all the waveguides are focusing. These conclusions are in good agreement with the results depicted in Figs. 2(a) and 2(b) and the fact that the effective diffraction of the array is anomalous in this case. Numerical solution associated with this case is depicted in Fig. 6(a). Meantime, bright solitons in the finite gap are expected to exist when PIM and NIM waveguides have nonlinearities of different signs, e.g., if $\chi_1 = -1$ and $\chi_2 = 1$ then the quantity $(\chi_1+\beta^4\chi_2)$ is negative for the top of the finite gap and positive (no solitons) for the bottom of the finite gap and other way round, when $\chi_1 = 1$ and $\chi_2 = -1$, despite the fact that the gap edges are symmetric (this feature is a reminiscence of the inversion symmetry breaking in the reciprocal space, reported in [37]). If $C_2 > C_{2q=0}^{(1)zd}$ then $\gamma$ is positive and the shift of the soliton's eigenvalue should be towards the external semi-infinite band gaps. The quantity $(\chi_1+\beta^4\chi_2)$ should be positive and bright solitons exist in focusing array in this case which is consistent with the fact that the effective diffraction of the array is normal now [Figs. 2(c) and 2(d)].

*Corresponding author.
E-mail: dovgiyalexandr@gmail.com

The numerical results have shown that more then one soliton families bifurcate from the gap edges of the linear spectrum [Figs. 5(a) and 5(c)]. The deeper is the soliton's eigenvalue lies in the band gap, the more localized the solitons are [Fig. 5(e)] and the higher the corresponding power level is [Fig. 4]. But near the zero diffraction points in the array these soliton solutions can be highly localized states even with the eigenvalues closely located to the gap edge, hence, at low power levels [Fig. 5(a)].

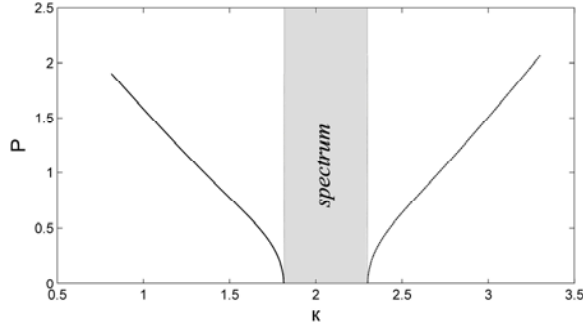

FIG. 4. Power $P$ vs $\kappa$ diagram of stable bright solitons in defocusing array at $q = 0$ (solid line) and in focusing array at $q = \pi$ (dashed line) when $\omega = 2.5$, $C_1 = 1$, $C_2 = 0.1$.

In Fig. 4 the power $P$ associated with bright soliton numerical solutions is depicted as a function of the eigenvalue $\kappa$ for the base and edge of the Brillouin zone. In both cases the behavior of the $P(\kappa)$ diagram is like that one in the standard discrete NLS model [9]. For relatively small values of power and weakly localized solutions, the behavior of the $P(\kappa)$ can be approximately described within the NLS limit. Using Eq. (2) and the expression (20) for the bright soliton solution with $q = 0$ in the continuous approximation, the $P(\kappa)$ curve can be approximately described by $P \approx 4(1+\beta^2)[\gamma(1-\beta^2)(\kappa - k^{(0)}_{q=0})]^{1/2} / |\chi_1 + \beta^4 \chi_2|$ at the base of the Brillouin zone. On the other hand, for high power levels and strongly localized solutions, most of the power is confined in one waveguide, and, therefore, $P \approx 2(1-\beta^4)(\kappa - k^{(0)}_{q=0})/(\chi_1 + \beta^4 \chi_2)$ for the solutions with $q = 0$. These two approximated dependences can be easily seen in Fig. 4. With the change of $C_2$ any peculiarities in the behavior of the $P(\kappa)$ diagram don't occur and these two approximations remain valid. Thus, the deeper is the soliton's eigenvalue lies in the band gap, the larger the relative difference $(\kappa - k^{(0)}_{q=0})$ is and the higher the power $P$ is.

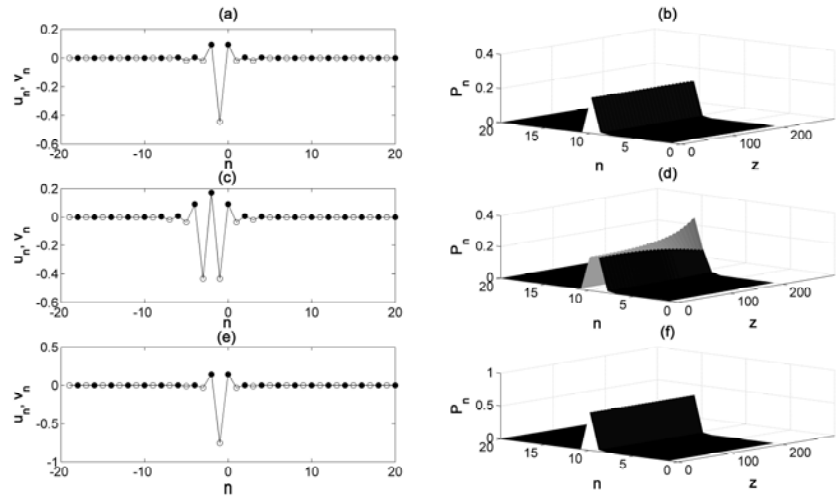

*Corresponding author.
E-mail: dovgiyalexandr@gmail.com

FIG. 5. Unstaggered discrete solitons with different eigenvalues within the positive infinite gap: (a) $\kappa = 2.3$, (c) $\kappa = 2.3$ (unstable excited state) and (e) $\kappa = 2.7$; and the corresponding power evolution depicted in (b), (d), (f), respectively, in the focusing array ($\chi_{1,2} = 1$) when $\omega = 2.5$, $C_1 = 1$, $C_2 = 0.2$. The infinite gaps are $|\kappa| > 2.1$. Empty (filled) circles indicate PIM (NIM) waveguides.

Now we will examine the stability of these bright soliton solutions with the linear stability analysis. We introduce perturbations in the exact solution $a_n = u_n \exp(i\kappa z)$ and $b_n = v_n \exp(i\kappa z)$ in a fashion $a_n = [u_n + (U_n + Q_n)\exp(i\Lambda z) + (U_n - Q_n)\exp(-i\Lambda z)]\exp(i\kappa z)$ and $b_n = [v_n + (V_n + W_n)\exp(i\Lambda z) + (V_n - W_n)\exp(-i\Lambda z)]\exp(i\kappa z)$, where $U_n$, $Q_n$, $V_n$ and $W_n$ are assumed to be small in comparison with $u_n$ and $v_n$. Substituting these perturbed solutions to Eq. (1) and linearizing it in small perturbations we arrive at the following coupled eigenvalue problem:

$$\begin{cases} -kQ_n + \omega Q_n + K_1(W_{n-1} + W_n) + K_2(Q_{n-1} + Q_{n+1}) + \chi_1 u_n^2 Q_n = \Lambda U_n, \\ -kU_n + \omega U_n + K_1(V_{n-1} + V_n) + K_2(U_{n-1} + U_{n+1}) + 3\chi_1 u_n^2 U_n = \Lambda Q_n, \\ -kW_n - \omega W_n - K_1(Q_{n+1} + Q_n) - K_2(W_{n-1} + W_{n+1}) - \chi_2 v_n^2 W_n = \Lambda V_n, \\ -kV_n - \omega V_n - K_1(U_{n+1} + U_n) - K_2(V_{n-1} + V_{n+1}) - 3\chi_2 v_n^2 V_n = \Lambda W_n. \end{cases}$$

When all eigenvalues $\Lambda$ are real the solution is stable, whereas, if an eigenvalue possesses a nonzero imaginary part the solution becomes unstable. We've analyzed this problem numerically. As can be seen from Figs. 5(a) and 5(c), solitons with different values of power $P$ exist in the array at the same parameters of the system. The explanation comes from Eq. (19). Indeed, the amplitudes $U_0$ and $V_0$ can be centered either at a PIM [Fig. 5(a)] or at a NIM [Fig. 5(c)] waveguide, thus giving two different families of soliton solutions which differ in the power level. The stability analysis has shown that the soliton solution with higher power level (excited state) is unstable [Fig. 5(d)]. The eigenvalue problem has complex roots $\Lambda$ with nonzero imaginary part that leads to an exponential growth of the amplitudes of the small perturbations on the soliton's background. The bright soliton solutions with lower power levels (ground state) are stable over the wide range of parameters of the system in question [Figs. 5(b), 5(f) and 6(b)]. All the eigenvalues $\Lambda$ are real, the instability doesn't occur and the perturbations are the small oscillations on the top of the soliton's background.

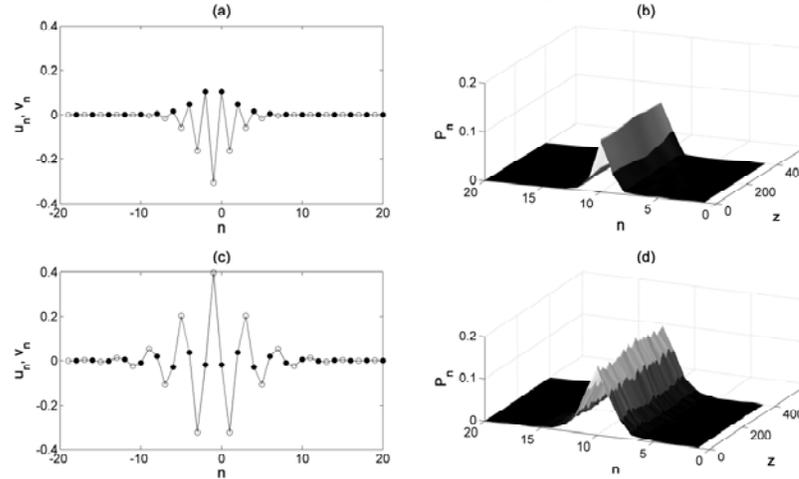

FIG. 6. Unstaggered (a) and staggered (c) discrete solitons in the finite gap with $\kappa = 1.9$ when $C_2 = 0.15$ and with $\kappa = 1.43$ when $C_2 = 0.5$, respectively, and the corresponding power evolution depicted in (b), (d), respectively, in the defocusing array ($\chi_{1,2} = -1$) when $\omega = 2.5$, $C_1 = 1$.

*Corresponding author.
E-mail: dovgiyalexandr@gmail.com

## B. At the edge of the Brillouin zone: $q=\pi$

Here we will study the properties of staggered soliton solutions, i.e., $q = \pi$. One can easily obtain the equations for these solutions from Eq. (18) by making the change of variables $u_n \to (-1)^n u_n$, $v_n \to (-1)^n v_n$ and hence, $\alpha \to -\alpha$. It's obviously that the relation which establishes domains of the solution's eigenvalue for which solitons exist doesn't change. Hence, suitable eigenvalues of staggered solitons lie in the band gaps too. To perform the multiple-scale analysis for this case we should also shift the soliton solution's eigenvalue towards the gaps. The only difference is that the gap edges should be taken at $q = \pi$. Thus, we consider $\kappa = k_{q=\pi}^{(0)} + \kappa_2 \varepsilon^2$, where $k_{q=\pi}^{(0)}$ indicate gap edges as were determined in the Sec. III at the edge of the Brillouin zone. The first-order term of the multiple-scale series gives that at the positive (negative) gap edge $V_0 = 0$, $U_0 \neq 0$ and $V_1 = [C_1/2(\omega - 2C_2)]dU_0/dx$ ($U_0 = 0, V_0 \neq 0$ and $U_1 = -[C_1/2(\omega - 2C_2)]dV_0/dx$), i.e., field amplitudes in PIM and NIM waveguides have different orders of magnitude [see Figs. 6(c), 7(a) and 7(c)]. Proceeding up to $\varepsilon^3$ we arrive at the stationary NLS equation for the positive (negative) gap edge:

$$\sigma \frac{d^2 U_0}{dx^2} - \kappa_2 U_0 + \chi_1 U_0^3 = 0 \quad (\sigma \frac{d^2 V_0}{dx^2} + \kappa_2 V_0 + \chi_2 V_0^3 = 0), \qquad (21)$$

where $\sigma = (4C_2^2 - 2\omega C_2 + C_1^2)/2(\omega - 2C_2)$ can be interpreted as the second-order diffraction coefficient by analogy with $\gamma$ and it's equal to zero when $C_2 = C_{2q=\pi}^{(1)zd}$. Note that in Eq. (21) the nonlinearity of only PIM (NIM) waveguides has influence due to the fact of different field magnitude's orders. The bright soliton solution of Eq. (21) is $U_0 = B/\cosh(ax)$ ($V_0 = B/\cosh(ax)$), where $B = [2\kappa_2/\chi_1]^{1/2}$ and $a = [\kappa_2/\sigma]^{1/2}$ ($B = [-2\kappa_2/\chi_2]^{1/2}$ and $a = [-\kappa_2/\sigma]^{1/2}$) for the positive (negative) gap edge. When $C_2 < C_{2q=\pi}^{(1)zd}$ the bright solitons bifurcate from gap edges towards the external semi-infinite gaps in focusing array [Figs. 7(a) and 7(c)] cause $\sigma$ is positive and the effective diffraction of the array is normal at the edge of the Brillouin zone [Figs. 2(a) and 2(b)]. Otherwise, no bright solitons can bifurcate from gap edges in focusing array when $C_2 > C_{2q=\pi}^{(1)zd}$.

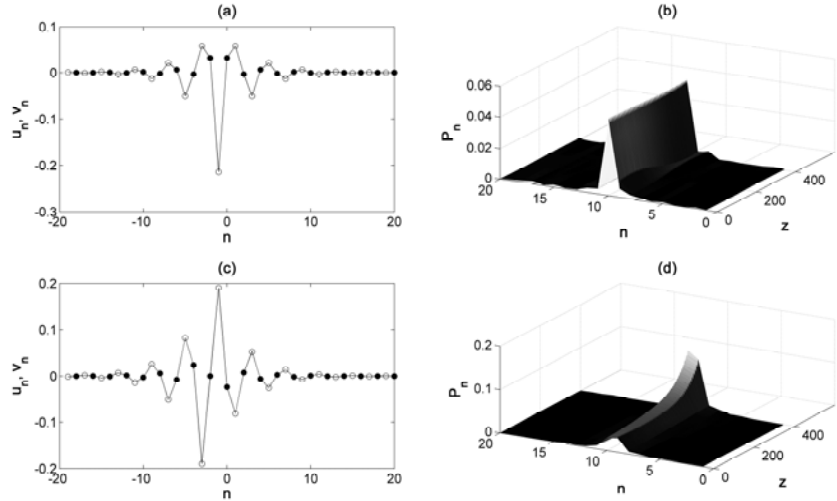

FIG. 7. Staggered discrete solitons in the positive infinite gap: (a) stable symmetric state and (c) unstable antisymmetric state; and the corresponding power evolution depicted in (b), (d), respectively, in the focusing array ($\chi_{1,2} = 1$) when $\omega = 2.5$, $C_1 = 1$, $C_2 = 0.19$, $\kappa = 2.14$. The infinite gaps are $|\kappa| > 2.12$.

*Corresponding author.
E-mail: dovgiyalexandr@gmail.com

In this case the effective diffraction of the array is anomalous [Fig. 2(d)], $\sigma$ is negative and the staggered bright solitons exist in defocusing array [Fig. 6(c)] with eigenvalues lying near the finite gap edges [Fig. 2(d)]. Using the analytical expressions for the soliton solutions of Eq. (21) bifurcating from positive (negative) gap edge at $q = \pi$ and Eq. (2), we can approximately describe the $P(\kappa)$ curve by $P = 4[(\kappa - k_{q=\pi}^{(0)})\sigma]^{1/2}/|\chi_1| + C_1^2(\kappa - k_{q=\pi}^{(0)})^{3/2}/3(\omega - 2C_2)^2|\chi_1|\sigma^{1/2}$ ($P = 4[(k_{q=\pi}^{(0)} - \kappa)\sigma]^{1/2}/|\chi_2| + C_1^2(k_{q=\pi}^{(0)} - \kappa)^{3/2}/3(\omega - 2C_2)^2|\chi_2|\sigma^{1/2}$) at low power levels. For strongly localized solutions at large values of power the $P(\kappa)$ curve can be approximately described by $P = 2(\kappa - k_{q=\pi}^{(0)})/\chi_1$ ($P = 2(k_{q=\pi}^{(0)} - \kappa)/\chi_2$) [Fig. 4]. With the deepening of the eigenvalue into the band gap the power increases and the staggered solitons become more localized.

The stability of these staggered soliton solutions is investigated similarly to the stability of unstaggered soliton solutions as was performed in the previous subsection. As can be seen from Figs. 7(a) and 7(c), symmetric and antisymmetric staggered soliton families exist in the array at the same parameters of the system. Numerical stability analysis has shown that this antisymmetric state is unstable [Fig. 7(d)]. The symmetric staggered solitons are stable both in focusing and in defocusing array [Figs. 7(b) and 6(d)]. The physical reason for this behavior is that the antisymmetric soliton is an excited state with higher power levels in comparison with symmetric soliton.

## V. CONCLUSIONS

In this paper we report on existence and properties of discrete gap solitons in binary nonlinear waveguide array of alternating positive and negative index waveguides with extended interactions. The zigzag geometrical configuration of the array in question allowing introducing extended strong second-order (next-to-nearest neighbors) couplings in addition to the first-order (nearest neighbors) coupling. The controllability of this second-order coupling allows managing the diffraction properties of this array. The effective diffraction can be controlled both in size and sign, it can be both normal and anomalous in the same system and even zero diffraction points exist their under definite values of second-order coupling coefficient. We've investigated modulation instability in focusing array and determined the regions where the continuous waves are stable or unstable as a function of the spatial Bloch momentum vector and the second-order coupling coefficient. Modulation instability doesn't occur in the regions of anomalous diffraction, whereas in the regions of normal diffraction the continuous waves are unstable both at the base and at the edge of the Brillouin zone. Due to the alternating positive and negative index waveguides the linear spectrum has a band gaps giving origin for more then one bright soliton families bifurcating from gap edges. The discrete solitons with the lowest power level are stable over a wide range of parameters. Discrete self-focusing is observed both in focusing and defocusing, and even in alternating focusing-defocusing array, moreover, near the zero diffraction points the highly localized states are possible at low power levels both for the base and for the edge of the Brillouin zone. Thus, the array considered is a more general model combining the properties of the arrays considered in Refs. [9] and [37], and provides more ways to manipulate energy localization effects, diffraction management and spatial discrete solitons formation.

**Acknowledgements**

We are grateful to A.I. Maimistov and E.I. Ostroukhova for useful and fruitful discussions. This work was supported by the Russian Scientists Found (project No. 14-22-00098).

*Corresponding author.
E-mail: dovgiyalexandr@gmail.com

*Corresponding author.
E-mail: dovgiyalexandr@gmail.com

*Corresponding author.
E-mail: dovgiyalexandr@gmail.com